\documentclass[1p,times,preprint]{elsarticle}
\usepackage[english]{babel}
\usepackage{tensor}
\usepackage{graphicx,psfrag}%,subfigure}
\usepackage{amsmath}
\usepackage{amssymb}
\usepackage{amsfonts}
\usepackage{dcolumn}
\usepackage{bm}
\usepackage{xcolor}
\usepackage{ulem}
\usepackage{tikz}
\usepackage{pgfplots}
\usepackage{subcaption}
\usepackage{comment}
\usepackage{verbatim}
\usepackage{fancyvrb}
\usepackage{cancel}

\def\ang#1{\langle #1 \rangle}

\def\bra<#1|{\mathinner{\langle\,{#1}\,\vert}} 
\def\ket|#1>{\mathinner{\vert\,{#1}\,\rangle}} 
\def\red|#1|{\mathinner{\!\vert\,{#1}\,\vert\!}}
\def\braket<#1>{\mathinner{\langle\,{#1}\,\rangle}} 
\begin{document}
%%%%%%%%%%%%%%%%%%%%%%%%%%%%%%%%%%%%%%%%%%%%%%%%%%%%%%%%%%%%%%%%%%%%%%%%%%%%%%%

\title{Relativistic variational methods and the Virial Theorem}
\date{\today}

\author[bcu]{Charlotte Froese Fischer}  
\address[bcu]{%
Department of Computer Science, University of British Columbia, 
%2366 Main Mall,
Vancouver, Canada
%BC V6T1Z4, 
}
\ead{cff@cs.ubc.ca}

\author[ulb]{Michel Godefroid}
\address[ulb]{%
Spectroscopy, Quantum Chemistry and Atmospheric Remote Sensing,
Universit\'e libre de Bruxelles,  
Brussels, Belgium
}
\ead{mrgodef@ulb.ac.be}

%%%%%%%%%%%%%%%%%%%%%%%%%%%%%%%%%%%%%%%%%%%%%%%%%%%%%%%%%%%%%%%%%%%%%%%%%%%%%%%
%
%   ABSTRACT
%
%%%%%%%%%%%%%%%%%%%%%%%%%%%%%%%%%%%%%%%%%%%%%%%%%%%%%%%%%%%%%%%%%%%%%%%%%%%%%%%
\begin{abstract}
In the case of the one-electron Dirac equation with a point nucleus the Virial Theorem (VT) states that the ratio of the kinetic energy to potential energy is exactly $-1$, a ratio that can be an independent test of the accuracy of a computed solution. This paper studies the virial theorem for subshells of equivalent electrons and their interactions in many-electron atoms. It shows that some Slater integrals impose conditions on a single subshell but others impose conditions between subshells.  The latter slow the rate of convergence of the self-consistent field process in which radial functions are updated one at a time. 
Several cases are considered.
\end{abstract}

\maketitle

\newpage

%%%%%%%%%%%%%%%%%%%%%%%%%%%%%%%%%%%%%%%%%%%%%%%%%%%%%%%%%%%%%%%%%%%%%%%%%%%%%%%

%%%%%%%%%%%%%%%%%%%%%%%%%%%%%%%%%%%%%%%%%%%%%%%%%%%%%%%%%%%%%%%%%%%%%%%%%%%%%%%
%
%   SECTION 1. INTRODUCTION
%
%%%%%%%%%%%%%%%%%%%%%%%%%%%%%%%%%%%%%%%%%%%%%%%%%%%%%%%%%%%%%%%%%%%%%%%%%%%%%%%

\section{Introduction}
\label{sec1}
For multi-electron atoms, the variational method for orbitals leads to a system of equations, one for each orbital $a$ with quantum numbers $n\kappa$, whose radial functions are varied.  In the numerical multiconfiguration Dirac-Hartree-Fock method (MCDHF), these equations have the  form: 
\begin{eqnarray}
 \label{eq:MCDHF}
 \left( \left[ \begin{array}{c c}
   V(r)   &  -c\left[ \frac{d}{dr} - \frac{\kappa}{r} \right] \\
      c\left[ \frac{d}{dr} + \frac{\kappa}{r} \right] & V(r)  -2c^2
      \end{array}\right]  
 \;  + \; \epsilon_aS \right) u_a(r) 
%      + \left(\frac{1}{r}\right)
\; + \;  \left[\begin{array}{c}
       X^{(1)}(r)\\
       X^{(2)}(r)
       \end{array}\right] = 0.
 \end{eqnarray}
 where $c$ is the constant used for the speed of light, $u_a$ a two-component vector of large,  $(P_a(r))$, and small $(Q_a(r))$ components of the radial functions, i.e., $u_a(r) = (P_a(r), Q_a(r))^t$,  
 $V(r)$ the potential, $\epsilon_a$ the orbital energy, and $S$ is a
 $2\times 2$ identity matrix in the $r$ variable. In numerical methods, for cases with two or more electrons,  the potential $V(r)$ includes only the direct interactions whereas all others are included in the two-component functions
 $X(r)$ along with contributions from Lagrange multipliers that assure orthogonality of orbitals. 
 The numerical solution of these equations involves the matching of outward and inward integration procedures.  For more details, see Ref.~\cite{review,grant,mcdf,FroSen:2020a}. 
 .  
 In B-spline methods
 all variational contributions to the energy expression are included in a matrix and orthogonality is imposed through projection operators
 ~\cite{dbsr-hf,bspline-advances}. The orbital energy $\epsilon_a$ is an eigenvalue of an interaction matrix. 

 The above equations represent a non-linear system of equations of eigenvalue type that are solved iteratively. Of interest is how accurately the equations have been 
 solved in the general case where the total energy of the system is a linear combination of one- and two-electron integrals. For each orbital that is varied, conditions relate the positive kinetic and negative potential energies. When all orbitals are varied, a virial-theorem condition may also be applied to the total energy. 
 
\begin{table}[h]
\caption{
Total energies,  $\ang{T}+\ang{V}$ (all in  $E_{\text{h}}$), and $\ang{V}/\ang{T}$ ratio
 for point-nucleus calculations of neutral atoms using the
$\mathtt{DBSR-HF}$~\cite{dbsr-hf} program
(see text for the grid and convergence parameters).
}
    \label{tab:neutrals}
    \centering
    \begin{tabular}{c l r r r } 
\hline
 Atom & Configuration             & \multicolumn{1}{c}{Total Energy $(E)$} & \multicolumn{1}{c}{$\ang{T} + \ang{V}$} 
                                                                     & \multicolumn{1}{c}{$\ang{V} / \ang{T}$} \\
 \hline
 He & $1s^2 $                     &    $-$2.861813342212 & $-$0.000000000118  & $-$1.000000000021 \\
 Be & [He]$2s^2$                  &   $-$14.575892266403 &  0.000000001419  & $-$0.999999999951  \\ 
 Ne & [Be]$2p^6$                  &  $-$128.691969446591 & $-$0.000000106271  & $-$1.000000000412  \\  
 Ar & [Ne]$3s^23p^6 $             &  $-$528.684450275764 &  0.000000632318  & $-$0.999999999404 \\
 Kr & [Ar]$3d^{10}4s^24p^6$       & $-$2788.884833711547 & $-$0.000006752662  & $-$1.000000001194 \\
 Rn & [Hg]$6p^6$                  &$-$23611.192499805627 & $-$0.000001513603  & $-$1.000000000029 \\
 \hline
 \end{tabular}
\end{table}

In order to better understand the problem, Table~\ref{tab:neutrals} reports the total energy ($E$), kinetic energy ( $\ang{T}$ ), potential energy ($ \ang{V}$), all in $E_{\text{h}}$ for the neutral atoms, He~$1s^2$, Be~[He]$2s^2$, Ne~[Be]$2p^6$, Ar~[Ne]$3s^23p^6$, Kr~[Ar]$3d^{10}4s^24p^6$, and Rn~[Xe]$4f^{14}5d^{10}6s^26p^6$ using the \Verb!DBSR-HF! program~\cite{dbsr-hf} to high precision.
To improve accuracy the first non-zero grid point was set to be $0.00001/Z$, the exponential grid-step parameter was reduced to {\tt he=0.125}, and the convergence for the energy was set to {\tt scf\_tol=1.d-16},  
 the change in the largest value of the radial function to {\tt orb\_tol=1.d-10}, and the tolerance for the tail region to {\tt end\_tol=1.d-10}. 
The first two tolerances were tolerances for a relative change. The entire calculation was done in double precision arithmetic with 15-16 digits of numerical accuracy. In other words the accuracy requested was machine precision for the total energy.
What we see immediately is that $\ang{T}+\ang{V}$ gets larger in magnitude as the number of shells increases as well as the total energy, except the value for Kr with the $3d^{10}$ subshell where the value is especially large.  
In atomic spectroscopy, what is important are the digits after the decimal point in that wavelengths depend on energy differences that usually are a fraction of the $E_h$ unit. Also reported by 
\Verb!DBSR-HF! is the ratio $\ang{V}/\ang{T}$. 

In this paper we analyze some simple cases that provide insight into the general problem and provide tests on the accuracy of solutions of Dirac-Hartree-Fock equations. We will see that the virial theorem relates to interactions between subshells of equivalent electrons.

 %%%%%%%%%%%%%%%%%%%%%%%%%%%%%%%%%%%%%%%%%%%%%%%%%%%%%%%%%%%%%%%%%%%%%%%%%%%%%%%
%
%   SECTION 2. GENERAL THEORY
%
%%%%%%%%%%%%%%%%%%%%%%%%%%%%%%%%%%%%%%%%%%%%%%%%%%%%%%%%%%%%%%%%%%%%%%%%%%%%%%%
\section{\label{sec2}General theory}

In classical mechanics, the virial theorem (VT) is well known~\cite{Goletal:2000a}. It relates the total kinetic energy of a stable system of discrete particles, bound by potential forces, with that of the total potential energy of the system.  The validity of this theorem for atoms was proved by Finkelstein~\cite{finkelstein} for the Schr\"odinger equation in 1928. Fock~\cite{fock,fockcoll}, who used the variational method to derive the Hartree-Fock method for atoms, was the first to derive the theorem for the Dirac equations using a "stretching method", also referred to as  a ``scaling''~\cite{Low:59a} or ``dilation'' procedure~\cite{grant}. In this analysis we will consider scaling as a perturbation of the many-electron wave function, analyze the effect of such a perturbation on each of the orbital equations, and relate the results to the virial theorem for the total energy.

In order to analyse the effect on the energy of a single electron in a potential from the scaling of the radial function, it is convenient to write Eq.~(\ref{eq:MCDHF}) in operator form, namely
\begin{equation}
\label{eq:oform}
\Big( T(r) + V(r) + M + \epsilon_a\; S \Big)\; u_a(r)=0,
\end{equation}
where 
\begin{eqnarray}
 T(r)  = & c\left[\begin{array}{c c}
                0 & 
 - \left( \frac{d}{dr} - \frac{\kappa}{r} \right)   \\
  \frac{d}{dr} + \frac{\kappa}{r} & 0  \end{array} \right],\;\; 
V(r) = \left[\begin{array}{c c}
                V(r) & 0 \\
                0    & V(r) \end{array}\right], \nonumber \\
                & & \\
&  M  =  -2c^2 \left[\begin{array}{c c}
                0    & 0 \\
                0    & 1 \end{array}\right], \;\; 
 S =  \left[\begin{array}{c c}
                1    & 0 \\
                0    & 1 \end{array}\right].  \nonumber   
\end{eqnarray}
Here $T(r)$ is the kinetic energy operator, $V(r)$ the potential energy operator, $M$ is a constant matrix referred to as the mass operator, whereas $S$ is the identity matrix. 

Consider the scaling perturbation, $r \rightarrow \lambda r$ 
of the radial function, so that Eq.~(\ref{eq:oform}) becomes
\begin{equation}
\label{eq:oform_2}
\Big( T(r) + V(r) + M + \epsilon_a(\lambda)\;  S \Big)\, u_a(\lambda r)=0, 
\end{equation}
where $\epsilon_a$ is now a function of $\lambda$. Before deriving an expression for the energy parameter, let us "de-scale" the entire radial equation by replacing $r$  by $r/\lambda$ so that the equation becomes
\begin{equation}
\label{eq:final_form}
\Big( T(r/\lambda) + V(r/\lambda) + M + \epsilon_a(\lambda)\; S \Big)\, u_a(r)=0. 
\end{equation}
Notice that, whereas the scaled radial function went back to its original function, the operators in the radial equation changed.    

In Eq.~(\ref{eq:final_form}), the function $u_a(r)$ is normalized and so, multiplying on the left by $u^t_a(r)$ and integrating, it follows that
\begin{equation}
\label{eq:energy_function}
-\epsilon_a(\lambda) = \int_0^\infty u^t_a(r) \; \Big(\, T(r/\lambda)+V(r/\lambda)+ M\Big)\,u_a(r) dr
\end{equation}
For a  solution $u_a(r)$, stationary with respect to the perturbation parameter $\lambda$,  the energy  will satisfy the variational condition, 
\begin{equation}
\label{eq:variational-condition}
    \Big(\frac{d \, \epsilon(\lambda)}{d\lambda} \Big)_{\lambda=1} =  
   0.
\end{equation}

In order to continue the analysis, let us consider a continuous operator ${O}(r)$ and define the expectation value of the operator to be
\begin{equation}
\label{eq:Operator} 
    \ang{O}_a(\lambda)  = \int_0^\infty u^t_a(r) \,  O(r/\lambda) \,u_a(r)dr
\end{equation}
and  $\ang{O}_a \equiv \ang{O}_a(\lambda=1)$. 
Then, according to 
%Eq.~(\ref{eq:oform}),
Eqs.~(\ref{eq:oform_2},\ref{eq:energy_function}),
\begin{equation}
\label{eq:eform}
 \ang{T}_a + \ang{V}_a + \ang{M}_a + \epsilon_a =0, 
\end{equation}
and the variational condition of Eq.~(\ref{eq:variational-condition})
leads to 
\begin{equation}
\label{eq:var_cond_2}
  \Big(\frac{d \ang{T}_a(\lambda)}{d\lambda} + \frac{d \ang{V}_a(\lambda)}{d\lambda} \Big)_{\lambda=1}=0.
\end{equation}
Thus we have shown that the scaling of the energy depends only on the scaling of $T(r) + V(r)$. 

The scaling of $T(r)$ is simple.  
Clearly $ \kappa/r \rightarrow \lambda\,\kappa/r$ as $r\rightarrow r/\lambda$. The operator $ d/dr $ scales in the same manner,  so that 
\begin{equation}
    T(r/\lambda ) = \lambda\, T(r), \;\; \forall\; r. 
\end{equation} 
Hence
\begin{equation}
    \Big ( \frac{d \ang{T}_a(\lambda)}{d\lambda}\Big)_{\lambda=1} =  \ang{T}_a. 
\end{equation}
But in general, since differentiation with respect to $\lambda$ and integration with respect to $r$ commute, 
\begin{equation}
    \Big(\frac{ d\ang{O}_a(\lambda)}{d\lambda}\Big)_{\lambda=1}
    = \int_0^\infty u^t_a(r) \; \Big( \frac{\partial O(r/\lambda)}{\partial \lambda}\Big)_{\lambda=1}\;u_a(r)dr \; . 
\end{equation}
Let $\widetilde{r} = r/\lambda$. Then 
\begin{equation}
\label{eq:fock-rule}
    \Big( \frac{\partial O(r/\lambda)}{\partial \lambda}\Big)_{\lambda=1} = \Big(\frac{d O(\widetilde{r})}{d \widetilde{r}}\times \frac{d\widetilde{r}}{d\lambda}\Big)_{\lambda=1} = - r \frac{d}{dr}O(r). 
\end{equation}
The $-rd/dr$ operator was first introduced by Fock~\cite{fock,fockcoll} and we will refer to it as the Fock rule.  

The discussion so far has assumed a single electron in a potential. We will extend this to a single orbital in a Hartree potential for equivalent electrons in a given subshell, and finally, an orbital in a Hartree-Fock potential for multiple subshells of equivalent electrons, all of which are scaled.  We will show that, in general, the scaling condition of  Eq.~(\ref{eq:var_cond_2}) resulting from the stationary condition of the orbital energy  Eq.~(\ref{eq:variational-condition}) with respect to the variation of  $\lambda$ (at $\lambda=1)$),  has the form
 \begin{equation}
\label{stable_lambda}
 \ang{ T }_a  + \ang{V}_{a} + \ang{ W }_a  = \ang{R}_a \; ,
\end{equation}
where $\ang{R}_a$ is referred to as a residual for the orbital equation, a quantity that is zero for exact solutions in the case of a single orbital but may not be zero in the case where the variation includes multiple orbitals. At the same time, it is referred to as a residual because, in numerical work, results will not be exactly zero, in which case $\ang{R}_a$ reflects the  accuracy of the numerical procedures. 
 Expressions for $\ang{W}_a$ appearing in Eq.~(\ref{stable_lambda}) will be determined, as a correction arising from the non-hydrogenic potential.  We refer to this expression as the orbital virial theorem  (OVT) to distinguish it from  the VT for the total energy for systems with multiple orbitals and 
 will show that $\sum_a q_a \ang{R}_a = 0$   for a solution that satisfies the virial theorem.  Here $q_a$ is the occupation number of shell $a$.
 
%%%%%%%%%%%%%%%%%%%%%%%%%%%%%%%%%%%%%%%%%%%%%%%%%%%%%%%%%%%%%%%%%%%%%%%%%%%%%%%
%
%   SECTION 3. Point nucleus: $V(r) = -Z/r$
%
%%%%%%%%%%%%%%%%%%%%%%%%%%%%%%%%%%%%%%%%%%%%%%%%%%%%%%%%%%%%%%%%%%%%%%%%%%%%%%%
\section{\label{sec3} Point nucleus: $V(r) = -Z/r$}

In the special case of a single electron in a potential for an atom or ion with a point nucleus,  $V(r) = -Z/r$, where  $Z$ is a constant representing the nuclear charge. Hence, like the kinetic energy,  
$\ang{{V}}_{a}(\lambda)  = \lambda \ang{V}_a $. 
Taking the partial derivative with respect to $\lambda$, and then setting $\lambda =1$ and $R(r)=0$ for an  exact solution for a single electron, Fock~\cite{fock,fockcoll} obtained the following OVT
\begin{equation}
\label{vt1}
 \ang{ T }_a  +  \ang{ V }_a   = 0  \mbox{\ \ or\ \  }    \ang{ T }_a /\ang{ V }_a = -1 \; .
\end{equation}
Essentially, the kinetic energy  and the  potential energy are equal in magnitude but opposite in sign, with the kinetic energy being positive, and with the sum being a test of accuracy of the computed solution.  
In this case,  the potential function is said to be homogeneous of degree $-1$ (ie. scales as $\lambda$)~\cite{Kim:67a} and  $\ang{W}_a=0.$

Consider some examples.  The recently proposed integration method for the Dirac equation~\cite{FroSen:2020a},  based on Simpson's rules,  solves the hydrogenic equation for a point nucleus to almost machine precision for high $Z$ as shown in Table~\ref{hyd-pt}. There are two more significant digits for hydrogen-like fermium, $Z=100$, than for the neutral hydrogen atom which suggests that the sum represents an absolute error rather that a relative error.  

\begin{table}[h]
\caption{Comparison of the  kinetic $\ang{ T }_{1s}$ and potential 
$ \ang{ V }_{1s} $ energies for a $1s$ electron for $Z=1$ and $Z=100$ hydrogen-like atoms.  For an  exact solution, $\ang{ T }_{1s}  + \ang{ V }_{1s} = 0$ but computationally  $\ang{ T }_{1s}  + \ang{ V }_{1s} = \ang{R}_{1s}.$}
\label{hyd-pt}
    \centering
    \begin{tabular}{ c r r } 
    \hline
  Type     &  \multicolumn{1}{c}{$ Z=1 $ }  & \multicolumn{1}{c}{$ Z=100$}\\
  \hline
 $\ang{ T }_{1s}$               &    1.0000266267326230   &   14625.659855116470 \\
 $\ang{ V }_{1s}$               & $-$1.0000266267322626   &$-$14625.659855116446 \\
 $\ang{ R }_{1s}$               &  0.0000000000003604     &   00000.000000000024 \\
    \hline
    \end{tabular}
 \label{hyd-pt}
\end{table}

%%%%%%%%%%%%%%%%%%%%%%%%%%%%%%%%%%%%%%%%%%%%%%%%%%%%%%%%%%%%%%%%%%%%%%%%%%%%%%%
%
%   SECTION 4. General Case:  $V(r) = -Z(r)/r$
%
%%%%%%%%%%%%%%%%%%%%%%%%%%%%%%%%%%%%%%%%%%%%%%%%%%%%%%%%%%%%%%%%%%%%%%%%%%%%%%%
\section{\label{sec4}General Case:  $V(r) = -Z(r)/r$ }

In the case of multiple equivalent electrons, the effective nuclear charge varies as a function of $r$ as in the Hartree-potential for a subshell, or as in the case of a finite nucleus, where the potential near the nucleus is modified.  
For such a situation, where the potential can be represented as  $V(r) = -Z(r)/r$, the Fock rule yields
\begin{equation}
\label{eq:rdvdr_2}
 \left ( 
    \frac{ d \ang{V}_a(\lambda)}{{d} \lambda} 
\right)_{\lambda = 1}
= \left\langle -  r \frac{ d V(r)}{d r}  \right\rangle _a 
= \left\langle V(r) \right\rangle_a + 
\left\langle \frac{ d Z(r)}{d r} \right\rangle_a \; .
\end{equation} 
Thus scaling  of this potential has introduced a new type of term, namely $\ang{W}_a= \ang{d Z(r)/dr}_a = \ang{Z'(r)}_a$. Notice that it depends on a derivative. Therefore, it
is not affected by adding or subtracting a constant. If $Z(r) = Z -s(r)$, 
then  $\ang{W}_a = -\ang{s'(r)}_a$, where $s(r)/r$ represents the deviation 
of  $V(r)$ from the Coulomb potential $-Z/r$. 

Given these quantities, the Virial Theorem  of an electron in a potential  remains that of Eq.~(\ref{stable_lambda}) but now $\ang{W}_a$ will not be zero.
Adding and subtracting  $ \ang{ W }_{a}$  to the orbital  energy equation~(\ref{eq:eform}),
we get :
\begin{equation}
\ang{ T }_a + \ang{ V }_a + \ang{ W }_a  + 
 \ang{ M }_a  - \ang{ W }_a + \epsilon_a = 0 \; .
\end{equation}
Noting that by Eq.~(\ref{stable_lambda}), the sum of the first three terms is $\ang{R}_a$, 
we obtain the second orbital virial relationship, namely
\begin{equation} 
\label{vt4}
\ang{ M }_{a} + \ang{R}_a -  \ang{ W }_{a} +  \epsilon_a = 0.
\end{equation}

Let us now consider two typical examples of potentials for a single subshell. 

\subsection{Finite volume nucleus}

In the case of a finite nucleus, Matsuoka and Koga~\cite{matsuoka} took a slightly different point of view. Their interest was primarily in the effect of a finite nucleus on the virial theorem for the total energy of many-electron systems.

Traditionally, for a finite nucleus, the potential  is defined as 
\begin{eqnarray}
   V(r)  = & -Z_{nuc}(r),\ \ & r < R_N  \nonumber  \\
           & -Z/r   &        r  > R_N,  
\end{eqnarray}
 where $R_N$ is the nuclear radius.  Hence the function $ Z'(r)$ would be non-zero only in the region $r < R_N$ . Their formula for the correction to the virial theorem was equivalent to $\ang{W}_a= -\ang{s'(r)}$
 using the uniform model, integrating over $(0,R_N)$. 
  
In the case of a Fermi nucleus~\cite{ParMoh:92a}, the potential and low-order derivatives are continuous at $r=R_N$.  In particular,
$-s'(r) \rightarrow 0$ as $r \rightarrow R_N$ and differentiation and/or integration can be over the range $(0,\infty)$. This form will be useful when both one-electron and many-body effects are  included in the potential.

\subsection{ The $1s^2$ $^1S_0$ case} 

Like the hydrogenic case, the variational equation for this two equivalent electron system is homogeneous in that there is no exchange contribution in Eq.~(\ref{eq:MCDHF}). Unlike the finite-nucleus effect that modifies only part of the orbital range, the potential for the $1s$ orbital is modified over the entire range by a contribution from the Slater integral $F^0(1s,1s)$ in the energy expression, but the virial equations have exactly the same form. The related contribution to $\ang{W}_{1s}$ can be computed directly by differentiating the potential for the  $1s$ orbital. There is only one orbital, and in this case $\ang{R}_{1s} = 0$ to numerical accuracy. 

Table~\ref{tab:general} shows the various contributions to the energy and the VT equations for these two cases -- the $1s$ potential of hydrogen-like fermium, with a Fermi nucleus, and the helium-like fermium $1s^2$ case with a point nucleus. The very common Fermi nucleus~\cite{ParMoh:92a,Paretal:96a} used in \Verb!GRASP!~\cite{g18} was used in the first case since, unlike the uniform distribution, the potential has continuous derivatives. The results here  were  obtained using the recently developed numerical procedures~\cite{FroSen:2020a}. For completeness, we also report the total energy, 
$$E(1s^2) =-2 \; \epsilon_{1s} - F^0(1s,1s).$$

\begin{table}[] 
\caption{ Energy contributions (in $E_{\text{h}}$) to virial theorems for hydrogen-like  (1-electron, Fermi nucleus) and helium-like (2-electron, point nucleus) fermium ($Z=100, A=257$) ground states. Parameters for the Fermi nuclear charge distribution, $\rho (r) = \rho_0 / (1 + e^{(r-b)/a})$, were  $r_{\text{rms}} = 5.8756$~fm, $a =  0.523387555$~fm, $b = 7.170561722$~fm. All quantities are for a single $1s$ orbital.} 
    \label{tab:general}
    \centering
    \begin{tabular}{c  r r}
    \hline
    Type & \multicolumn{1}{c}{$1s$} & \multicolumn{1}{c}{$1s^2$}\\
    \hline
$ \ang{T}_{1s} $       &     14431.431739850053  &    14512.968453372794 \\
$ \ang{V}_{1s} $       &    $-$14453.622410065407  &   $-$14472.308294232836 \\
$ \ang{M}_{1s} $       &     $-$5900.425512587179  &    $-$5898.429859105305 \\
$ \ang{W}_{1s}  $      &        22.190670173520  &      $-$40.660159132020 \\
$\epsilon_{1s}$        &      5922.616182802533  &     5857.769699965348  \\
$E(1s^2)$              &                         &   $-$11796.859719783146 \\    
$ \ang{R}_{1s}$ &      0.000000000001     &        0.000000000001 \\
$ \ang{M}_{1s} - \ang{W}_{1s} + \epsilon_{1s} $ &    0.000000000000  &       $-$0.000000000000 \\
  \hline
    \end{tabular}
\end{table}
These two cases are important test cases for any code of relativistic, computational, atomic structures.  There is only one orbital and the energy expression includes no exchange contributions. We see that, for a system with one orbital, $\ang{R}_a = 0$ for an exact solution.

%%%%%%%%%%%%%%%%%%%%%%%%%%%%%%%%%%%%%%%%%%%%%%%%%%%%%%%%%%%%%%%%%%%%%%%%%%%%%%%
%
%   SECTION 5. Many-electron DHF equations
%
%%%%%%%%%%%%%%%%%%%%%%%%%%%%%%%%%%%%%%%%%%%%%%%%%%%%%%%%%%%%%%%%%%%%%%%%%%%%%%%
\section{\label{sec5} DHF equations for multiple orbitals}

The theory presented so far has been discussed strictly as an electron ($a$) in a potential but it can readily be extended to a subshell (say $\alpha$) of $q_a$ equivalent electrons by multiplying the orbital properties by the occupation number.  For example, for the  $1s^2$ subshell containing two $1s$ electrons, we have
\begin{equation}
\label{eq:subshell_from_orbital}
\ang{T}_{1s^2} = 2 \ang{T}_{1s}, \ \ \ang{V}_{1s^2} = 2 \ang{V}_{1s}, \ \ 
\ang{W}_{1s^2} = 2 \ang{W}_{1s}, \ \ \ang{R}_{1s^2} = 2 \ang{R}_{1s}. 
\end{equation}
Multiplying also Eq.~(\ref{stable_lambda}) by $q_{1s}=2$, we get the OVT for the $1s^2$ subshell, namely
\begin{equation}
    \ang{T}_{1s^2} + \ang{V}_{1s^2} + \ang{W}_{1s^2} = \ang{R}_{1s^2}. 
\end{equation}
In this form, the radial equations for the orbitals are derived from the variation of a specific orbital without dividing by the occupation number for the subshell, as often is done for the derivation of the Hartree-Fock equations for multi-subshell atoms.

In order to derive expressions for the various kinetic, potential, and other properties of subshells, we will outline an algorithm for determining contributions from radial integrals and their coefficients as they appear in an energy expression and then derive the scaling equation for the total energy by summing over the subshells. This approach will include both direct and exchange contributions to the potential.  For an exact (self-consistent) solution, for which $\sum_a  q_a \ang{R}_a=0$,  we now have 
$\sum_\alpha \ang{R}_\alpha =0$ by Eq.~(\ref{eq:subshell_from_orbital}). 

In our analysis here we will consider Dirac-Hartree-Fock energy expressions that include only one-electron integrals, 
$$I(a,a) = \ang{T}_{a} +\ang{V_1}_{a} + \ang{M}_{a}$$
and  Slater integrals, 
$F^k(a,b)$ or $G^k(a,b)$ that are 2-electron contributions to $\ang{V_2}_\alpha$  where
$$\ang{V}_\alpha = \ang{V_1}_\alpha + \ang{V_2}_\alpha.$$ 
 Given an orbital basis and an energy expression, the kinetic and potential energies for subshells can readily be computed directly without considering orbital potentials. 
 
 The one-electron integrals $I(a,a)$, multiplied by their coefficients, contribute directly to $\ang{T}_\alpha$, $\ang{V}_\alpha$, and $\ang{M}_\alpha$, but contributions from Slater integrals that may depend on two different orbitals (subshells) and then contribute to both subshells, need further investigation.
 
\subsection{Slater integrals}

The relativistic Slater integral, usually denoted as $R^k(ab,cd)$, is a two-dimensional integral with two co-ordinates, say $(r, s)$. Let us denote radial factors in terms of vector products of large and small components
$$ \rho(ac;r) = u_a^t(r)u_c(r), \mbox{ and } \rho(bd;s) = u_b^t(s) u_d(s). $$
Slater integrals have many symmetries, arising from the symmetry of a product, and the symmetry of the co-ordinate system.  In the canonical form, orbitals are in a designated order such that $a <= c, b <= d$, $a <=b$ and, if $a=b$, then also $c <= d$. Consequently, a typical canonical integral is
\begin{equation}
\label{slater}
    R^k(ab,cd) = \int_0^\infty \left( \rho(ac;r)\frac{ Y^k(bd;r)}{r} \right) dr\ \ 
\equiv \int_0^\infty \left(\frac{Y^k(ac;r)}{r} \rho(bd;r)\right) dr \; ,
\end{equation}
where  
 \begin{eqnarray}
\label{eq:yk-def}
     Y^k(bd;r) &= &\int_0^r \left(\frac{s}{r}\right)^k \rho(bd;s)ds 
   + \int_r^\infty \left(\frac{r}{s}\right)^{(k+1)}\rho(b d ;s) ds  
\end{eqnarray}
 Notice that $Y^k(bd;r)$ has two contributions, and could be defined as  
\begin{eqnarray}
\label{eq:yk-def_2}
Y^k(bd;r)  & =  & r^{-k} \; A(bd;r) +   r^{k+1} \; B(bd;r) \;, 
\end{eqnarray}
with $A(bd;r)$ from charge density $s <r$, and $B(bd;r)$, from charge density $s >r$, namely
\begin{equation}
\label{eq:A_and_B_def}
A(bd;r) =  \int_0^r s^k \rho(b d ;s) ds \; \;\mbox{and} \;\;
B(bd;r) =  \int_r^\infty s^{-(k+1)}\rho(b d ;s) ds \; .
\end{equation}
Because the variable $r$ simply defines the range of integration, the derivatives of $A$ and $B$ obey a useful relation
\begin{equation}
\label{eq:deriv}
       r^{-k} A'(bd;r)  + r^{k+1} B'(bd;r) = 0 \; .
\end{equation}

The calculation of Slater integrals (and related $Y^k$ functions) is the most time-consuming aspect of atomic structure calculations. Hartree~\cite{Har:57a}, who did calculations with a desk calculator, found that the most efficient method was to solve two first-order differential equations that do not depend on powers of $r^k$ . The solution of the first differential equation for outward integration, was the function
\begin{equation}
\label{eq:zk-def}
    Z^k (bd;r) = r^{-k} \int_0^r s^k \rho(bd;s)ds,  
\end{equation}
whereas the second was inward integration of an equation for the function $Y^k(bd;r)$,  with the initial starting condition, $Y^k(bd;r_{max}) = Z^k(bd;r_{max})$. Procedures for  
solving these equations have recently been evaluated~\cite{FroSen:2020a}.

For the variation of the potential with respect to scaling we need to consider the contribution to a potential, say $V^k(bd;r) \equiv  Y^k(bd;r)/r$ or
\begin{equation}
\label{eq:Vk}
    V^k(bd;r) =  
     r^{-(k+1)} A(bd;r)  + r^{k} B(bd;r) \; .
\end{equation}

\subsection{Scaling of the $V^k(bd;r)$ operator}

Applying the Fock rule~(\ref{eq:fock-rule}) on the potential term $ V^k(bd;r)$ and using the property~(\ref{eq:deriv}), it readily follows that
\begin{equation}
\label{eq:fock-rule_on_Vk}
    \Big( \frac{\partial  V^k(bd;r/\lambda)}{\partial \lambda}\Big)_{\lambda=1}  = - r \frac{d}{dr}V^k(bd;r). 
\end{equation}
can be written as
\begin{eqnarray}
\label{eq:fock-rule_on_Vk_2}
\Big( \frac{\partial  V^k(bd;r/\lambda)}{\partial \lambda}\Big)_{\lambda=1}  & = & V^k(bd;r) + W^k(bd;r) = 
Y^k(bd;r)/r + W^k(bd;r) \nonumber \\
& = & (k+1)r^{-(k+1)}A(bd;r) -k r^k B(bd;r) \; .
\end{eqnarray}

 By definition~(\ref{eq:zk-def}),  $r^{-(k+1)}A(bd;r) = Z^k(bd;r)/r$. Using Eq.~(\ref{eq:yk-def_2}) it follows that  $r^k B(bd;r) = Y^k(bd;r)/r-Z^k(bd;r)/r$. 
Consequently, $W^k(bd;r) $ can be extracted from Eq.~(\ref{eq:fock-rule_on_Vk_2}) as 
 \begin{eqnarray}
     \label{eq:wk-def}
     W^k(bd;r) & = & Z^k(bd;r)/r -Y^k(bd;r)/r  + k\left( 2 Z^k(bd;r)/r - Y^k(bd;r)/r \right) \nonumber \\
& = & (2k+1)\; Z^k(bd;r)/r - (k+1) Y^k(bd;r)/r   
 \end{eqnarray}

\subsection{Symmetry of Slater integrals in DHF energy expressions } 

The above derivation was a general derivation for any combination of orbitals. In DHF calculations only $R^k(ab,ab)$ or $R^k(ab,ba)$ symmetries occur. Then, if $d=b$ and $c=a$
the contribution to the scaling   
$\ang{ W }_\alpha$ would be
\begin{equation}
   \ang{W^k(bb;r)}_{aa} = \int_0^\infty \rho(aa;r)W^k(bb;r) dr. 
\end{equation}
Here we have introduced the subscript $aa$ to denote the $\rho(r)$ function required for the calculation of the expectation value.  
Scaling in the other co-ordinate, yields a contribution to $ \ang{ W }_\beta$ as 
\begin{equation}
  \ang{W^k(aa;r)}_{bb} = \int_0^\infty \rho(bb;r)W^k(aa;r) dr. 
\end{equation}
In the case of a $G^k(ab)$ Slater integral there is only one possibility, namely  $Y^k(ab;r)$ and $\rho(ab;r)$,
and the integral
\begin{equation}
  \ang{W^k(ab;r)}_{ab} = \int_0^\infty \rho(ab;r) W^k(ab;r) dr,
\end{equation}
would contribute to both 
$ \ang{ W }_\alpha$ and $\ang{ W }_\beta$, multiplied by the coefficient of the Slater integral. 

\subsection{Total energy virial equation}  

Computationally, the list of one-electron integrals and Slater integrals, along with their coefficients, defines the total energy of a state and also 
determines the subshell quantities $\ang{T}_\alpha$, $\ang{V}_\alpha$, $\ang{M}_\alpha$, and $\ang{W}_\alpha$.  Then
\begin{equation}
\label{eq:total_cont}
  \ang{T} = \sum_\alpha \ang{T}_\alpha ,  \quad   \ang{M} = \sum_\alpha \ang{M}_\alpha ,  \quad  \ang{W} = \sum_\alpha \ang{W}_\alpha, \quad \ang{R} =  \sum_\alpha \ang{R}_\alpha
\end{equation}
are values that contribute directly to the total energy and its dilation equation.  However, the sum of potential energies of subshells is different.  Because the Slater integrals with orbitals $a,b$ from subshells $\alpha, \beta$
make contributions to both $\ang{V}_\alpha$ and $\ang{V}_\beta$ (or twice to $\ang{V}_\alpha $ when $b=a$), we have
\begin{equation}
  \sum_\alpha  
  \ang{V}_\alpha =  \ang{V} + \ang{V_2} 
 \end{equation}
 where $\ang{V}$ is the potential energy for the total energy and $\ang{V_2}$ represents the contribution from the 2-body Slater integrals.  Then we have the two equations, 
\begin{eqnarray}
\label{eq:u-def}
 \ang{T} + \ang{V} + \ang{M} &=& E \; ,  \nonumber \\
 \ang{T} + \ang{V} + \ang{U} & =& \ang{R} \quad \mbox{where} \quad \ang{U} = \ang{V_2} + \ang{W}.
\end{eqnarray}
Consequently,  $\ang{T} + \ang{V} = \ang{R} - \ang{U}$.  Substituting  into  the energy equation, we  get the mass dilation condition
\begin{equation}
\label{eq:e-def}
  \ang{M} + \ang{R} - \ang{U}  = E.  
\end{equation}
Thus an exact solution requires that $\ang{U}=0 $ and also $\ang{R}=0$ in which case it satisfies two conditions, namely $\ang{T} + \ang{V} =0$ and $ \ang{M} = E.$
%%%%%%%%%%%%%%%%%%%%%%%%%%%%%%%%%%%%%%%%%%%%%%%%%%%%%%%%%%%%%%%%%%%%%%%%%%%%%%%
%
%   SECTION 6. Results and analysis
%
%%%%%%%%%%%%%%%%%%%%%%%%%%%%%%%%%%%%%%%%%%%%%%%%%%%%%%%%%%%%%%%%%%%%%%%%%%%%%%%
\section{\label{sec6}Results and analysis}

 \begin{table}[] 
\caption{Trends in dilation data (all in  $E_{\text{h}}$) for the ground state of ions of Fm ($Z=100)$ with increasing number of closed shells obtained using 
$\mathtt{DBSR-HF}$ (Ref.~(\cite{dbsr-hf})) or $\mathtt{GRASP}$~(Ref.(\cite{g18})).   
Reported are the residuals for subshell equations,
$ \ang{ R}_\alpha = \ang{ T + V + W}_\alpha $,
the dilation correction for the total energy $\ang{U}$, and $\ang{R}$, the residual of the dilation equation for the total energy.
See text for details.}
%small
\scriptsize
\label{tab:fm-ions}
    \centering
    \begin{tabular}{l r r r r r r}
    \hline
      & \multicolumn{1}{c}{$ \ang{T }_\alpha $}
      & \multicolumn{1}{c}{$ \ang{V }_\alpha $}
      & \multicolumn{1}{c}{$ \ang{W }_\alpha $}
      & \multicolumn{1}{c}{$ \ang{R }_\alpha $}
      & \multicolumn{1}{c}{$\ang{U}$}
     & \multicolumn{1}{c}{$\ang{R}$}\\
    \hline
\multicolumn{3}{l}{He-like fermium, Ref.~(\cite{dbsr-hf})}&  &                             & 0.0000015974 & -0.0001437723 \\
  $1s$  & 29025.9667069950 & $-$28944.6465327985  &  $-$81.3203179688 &  $-$0.00014377231 & \\
  &&&&&& \\
\multicolumn{3}{l}{Be-like fermium, Ref.~(\cite{dbsr-hf})}& &                              &$-$0.0000004704 & $-$0.0001861707 \\
   $1s$ & 29002.8729873817 & $-$28825.0962333553  & $-$169.6017382566 &   8.17501576976 & \\
   $2s$ &  7639.5438558036 &  $-$7611.0901830102  &  $-$36.6288747339 &  $-$8.17520194041 & \\
  &&&&&& \\
\multicolumn{3}{l}{Ne-like fermium, Ref.~(\cite{dbsr-hf})}&  &                           &$-$0.0066586823 & $-$0.0046480020 \\
  $1s$ & 28887.5200623142 & $-$28428.3210384095  & $-$468.7333821954 &  $-$9.53435829067 & \\
  $2s$ &  7355.6539876266 &  $-$7224.0189346195  & $-$130.1606742682 &   1.47437873888 & \\
  $2p-$&  7279.6944610751 &  $-$7093.9299579689  & $-$187.5822025200 &  $-$1.81769941386 & \\
  $2p$ &  9758.9822408182 &  $-$9511.5495258556  & $-$237.5596839990 &   9.87303096364 & \\
 &&&&&& \\
 \multicolumn{3}{l}{Ar-like fermium, Ref.(\cite{g18})}&  &                                 & $-$0.0082541861 & $-$0.0058322935 \\
  $1s$ & 28853.8688681747  & $-$28228.5467404879  & $-$636.52909212681 & $-$11.2069644401 & \\
  $2s$ &  7288.4050654392  &  $-$7033.9196432434  & $-$249.87424551899 &   4.6111766768 & \\
  $2p-$&  7215.9989758875  &  $-$6898.1971349928  & $-$315.58078033595 &   2.2210605587 & \\
  $2p$ &  9641.7441568130  &  $-$9154.8935703478  & $-$473.15022927091 &  13.7003571942 & \\
  $3s$ &  2646.6802467869  &  $-$2572.8874558901  &  $-$77.91413433931 &  $-$4.1213434425 & \\
  $3p-$&  2604.0632334467  &  $-$2513.5288661661  &  $-$95.15214537503 &  $-$4.6177780945 & \\
  $ 3p$&  3932.1056407442  &  $-$3792.1568515901  & $-$140.54112990026 &  $-$0.5923407462 & \\
\hline
% \multicolumn{5}{l}{* Estimates deduced from DBSR-HF values of $\ang{T}+\ang{V}= -U$ (estimated).}
    \end{tabular}
\end{table}

% E(Ar) = -26143.684487736813  (DBSR_HF)  E=   -26143.681784528952 (G18) Diff= .002703207861

% we could save space by reporting fewer decimal places.
% end of fm-table file 

In order to illustrate the concepts that have been developed, Table~\ref{tab:fm-ions} shows results for the ground states of ions of Fm ($Z=100$) as more closed shells are added.  Included are values for 
He-like, Be-like, Ne-like studies using radial functions from \Verb!DBSR-HF!~\cite{dbsr-hf} for standard (not high-precision) calculations and Ar-like with radial functions from \Verb!GRASP!~\cite{g18}.  In each case, radial functions were transformed to the numerical grid of the revised procedures~\cite{FroSen:2020a} and the various components computed and analysed. 
Displayed is information about the scaling of  the individual subshells, the most important being $\ang{R}_\alpha$. Notice that for He-like, the value is small although somewhat larger than the value reported earlier in Table~\ref{tab:general} because the present values are not high-precision results. What is interesting is how the different $\ang{R}_\alpha$ values appear to be related and cancel when summed. In particular, in the Be-like case,  notice how $\ang{R}_{1s^2}$ cancels $\ang{R}_{2s^2}$ to yield a small $\ang{R}$. Intuitively, if the $1s$ orbital is too contracted it screens the nucleus more than it would for the exact solution, and the $2s$ orbital will be too expanded, or vice versa.  However, in scaling, both orbitals are either contracted ($\lambda < 1$) or expanded ($\lambda > 1$).
 
 In MCDHF calculations, the contribution to $\ang{U}$  will be zero if the contribution from each Slater integral is zero, namely that for each Slater integral its contribution to $\ang{U} = \ang{W} + \ang{V_2}$
is zero.  

Three cases need to be considered all with $k=0$, in our present examples.
 \begin{enumerate}
     \item $F^k(a,a)$:~~   
     By Eq.~(\ref{eq:wk-def}) the scaling is the same in both co-ordinates so  
     \begin{equation}
         2 \ang{W^0(aa;r)}_{aa} +F^0(a,a) = 2 \ang{Z^0(aa;r)}_{aa} - F^0(a,a) =0. 
     \end{equation}
     \item $G^0(a,b)$:~~
     In this case the contribution is the same for each co-ordinate and the sum (involving both subshells) is
     \begin{equation}
         2 \ang{W^0(ab;r)}_{ab} +G^0(a,b) = 2 \ang{Z^0(ab;r)}_{ab} - G^0(a,b) =0.
     \end{equation}
    \item  $F^0(a,b)$:~~
    In this case, the scaling depends on the co-ordinate and the effect of scaling for orbital $a$ differs from that for orbital $b$. However, the combined contribution to the dilation equation for the total energy should still be zero, namely
    \begin{equation}
    \ang{W^0(bb;r)}_{aa}  + \ang{W^0(aa;r)}_{bb} +F^0(a,b) =
      \ang{Z^0(bb;r)}_{aa} + \ang{Z^0(aa;r)}_{bb} - F^0(a,b) = 0.
    \end{equation}
 \end{enumerate}
 Similar equations hold for $k > 0$ since, as seen in Eq.~(\ref{eq:wk-def}), $k$ multiplies a factor
 \begin{equation}
     \left( 2 Z^k(bd;r)/r - Y^k(bd;r)/r \right)     
 \end{equation}
for which an expectation value should  be zero. 
 For example,  the list of integrals $F^k(3d,3d), k=0,2,4$ in the energy expression for a $3d^{10}$ subshell would have the conditions, $2\ang{Z^k(3d3d;r)}_{3d3d} - F^k(3d,3d) =0,$  associated with them for each $k$, a condition that might be difficult to satisfy exactly.
 By Eqs.~(\ref{eq:yk-def_2},\ref{eq:zk-def}), this condition, in general, reduces to 
 \begin{equation}
     \left( 2 Z^k(bd;r)/r - Y^k(bd;r)/r \right)  =  r^{-k} A(bd;r)  -  r^{k+1} B(bd;r). 
 \end{equation}
 Thus the virial theorem requires that certain expectation values of this difference be zero. In other words,
 the inner and outer contributions to the potential from the $\rho(bd;r)$ charge densities must be equal in magnitude, at least in the case of $F^k(a,a)$ and $G^k(a,b)$.  In the case of $F^k(a,b)$, two potentials and charge densities are involved, which probably explains why the $\ang{R}_\alpha$ values in Table~\ref{tab:neutrals} tend to increase. 

Let us confirm some results for the simple case of $1s^2$.  The formula for the energy is
$$E = 2 \, \Big(  \ang{T}_{1s} + \ang{V_1}_{1s} + \ang{M}_{1s} \Big)  + F^0(1s,1s).$$ 
Then the subshell equations for the energy and dilation are, respectively, 
\begin{eqnarray} 
 2 \ang{T}_{1s} + 2 \ang{V_1}_{1s} + 2 F^0(1s,1s) + 2 \ang{M}_{1s}  + 2 \epsilon(1s) = 0, \nonumber \\
 2 \ang{T}_{1s} + 2 \ang{V_1}_{1s} + 2 F^0(1s,1s)  +2 \ang{W}_{1s} = 2\ang{R}_{1s}.
\end{eqnarray} 
The latter may be written as
\begin{equation}
\label{eq:he-like}
 \ang{T} + \ang{V} + \ang{U} = \ang{R},
\end{equation} 
where 
$$\ang{V} = 2 \ang{V_1} + F^0(1s,1s)$$
and 
%\cff{Comment: some corrections here -- $1s^2$ rather than $1s$}
$$\ang{U} = F^0(1s,1s) +\ang{W}_{1s^2}=  F^0(1s,1s) + 2\ang{W}_{1s} $$
with 
$$ \ang{W}_{1s^2}  
= 2 \ang{ W^0(1s,1s;r)}_{1s1s} 
= 2 \ang{ Z^0(1s,1s;r)/r - Y^0(1s,1s;r)/r}_{1s1s} $$
by Eq.~(\ref{eq:wk-def}). 
As a result,  for an exact solution of $1s^2$, where $\ang{U} =0$ and $\ang{R}=0 $, we would have the condition
\begin{equation}
 \ang{T}/\ang{V} = -1  \quad  {\mbox and} \quad  \ang{Z^0(1s,1s;r)/r}_{1s1s} = 
 \frac{1}{2}\ang{Y^0(1s,1s;r)}_{1s1s} \equiv \frac{1}{2}F^0(1s,1s).
 \end{equation}
 Here $\rho(1s,1s;r)$ is both a weighting factor for the expectation value and the charge density function that determines a contribution to the potential as a function of $r$. 
 
\begin{table}[]
\caption{Comparison of the virial theorem analysis for a fully variational calculation for Be $1s^22s^2$ and a fixed core calculation where the $1s$ orbital is from the ion, Be$^{+2}$ $1s^2$.  }
    \label{tab:be-data}
    \centering
    \begin{tabular}{l r r r r } 
 \hline
\multicolumn{3}{l}{i) Fully Variational} & \multicolumn{2}{l}{$E = -14.5758922675 $}   \\
\hline
$\alpha$ &  $\ang{T}_\alpha$  & $\ang{V}_\alpha$ &  $\ang{W}_\alpha$ &  $\ang{R}_\alpha $  \\
\cline{2-5}
$1s^2$ &  27.1522686546       & $-$23.0458277992   & $-$4.0523847421     &   0.0540561132 \\
$2s^2$ &   2.0052562788       &  $-$1.6214451720   & $-$0.4378672909     &  $-$0.0540561841 \\
       &   $\ang{T} $         &  $\ang{V}$       &   $\ang{U}$       &  $ \ang{R}$   \\
       \cline{2-5}
$1s^22s^2$ &  29.1575249333   & $-$29.1575250573   & $-$0.0000000531     &  $-$0.0000000709 \\
\hline  \\
\multicolumn{3}{l}{ii) Fixed $1s$ from Be$^{+2}$ $1s^2$ core.} &
\multicolumn{2}{l}{$E = -14.5758644561$}   \\
\hline
$\alpha$ & $\ang{T}_\alpha$  & $\ang{V}_\alpha $ & $\ang{W}_\alpha$  &  $\ang{R}_\alpha $  \\
\cline{2-5}
$1s^2$ &   27.2334088761     &   $-$23.0833644426  & $-$4.0568852083     &  0.0931592252 \\
$2s^2$ &    1.9893037589     &    $-$1.6134422475  & $-$0.4364030288     & $-$0.0605415172  \\
       &   $\ang{T} $        &      $\ang{V}$     &   $\ang{U}$        & $ \ang{R}$  \\
  \cline{2-5}
$1s^22s^2$ & 29.2227126349   & $-$29.1900949804    &  0.0000000535     &  0.0326177080 \\
\hline 
 \end{tabular}
\end{table}

Another interesting case is displayed in Table~\ref{tab:be-data} for fully variational calculations of 
Be $1s^22s^2$. Notice the large values for $\ang{R}_{1s^2}$ and $\ang{R}_{2s^2}$ that are almost equal in magnitude but opposite in sign so the that virial condition for the configuration of the total energy is near zero. 
Frequently, when spectrum calculations are performed for multiple levels, a fixed core approximation is used.
Also shown in this table are the virial results for the case where the $1s$ orbital is fixed at the value for 
Be$^{+2}$ $1s^2$ and only the $2s$ orbital is varied. In this fixed core approximation, the $\ang{R}_\alpha$ values no longer cancel. The fixed core approximation is often desirable in the case of heavy elements. Possibly codes like \Verb!DBSR-HF! should report the virial theorem results only for the orbitals that are varied which means that the interaction of the varied subshells with the fixed core subshells needs to be omitted in the analysis.

Some spline methods for non-relativistic variational equations, updating a few  or all orbitals simultaneously, have been  investigated by Froese Fischer {\it et al.}~\cite{FroGuoShe:92}. It is a quadratically convergent Newton-Raphson (NR) method that achieves better performance.  In their study, the calculation for Be $1s^22s^2$ converged in four (4) iterations with a virial theorem deviating from $-2$ by $10^{-13}$, considerably fewer than the twelve (12) SCF  
iterations shown in Table~\ref{tab:conv} for the relativistic equation.
Newton-Raphson methods have not yet been applied to relativistic equations but are expected to perform in a similar manner.

 \begin{table}[] 
\caption{Comparison of the convergence of self-consistent (SCF) iterations for atoms Be ($Z=4$) and Rn ($Z=86$) and a point nucleus. For each iteration, (i),  the relative change in the total energy 
($\Delta_{\mbox{scf}}$)  and the relative change in the maximum value of the radial function ($\Delta_{\mbox{orb}}$) 
are reported, where  x.xxE$-$yy represents x.xx$\;\times\; 10^{-\mathtt{ yy}}$.}
\label{tab:conv}
    \centering
    \begin{tabular}{r r r r r r r }
    \hline \\
\ \ \ \  & \multicolumn{2}{c}{Be $1s^22s^2$ } & \ \  & \multicolumn{2}{c}{ Rn  [Hg]$6p^6$ } \\ \\
\cline{2-3} \cline{5-6}
\mbox{i} & $\Delta_{\mbox{scf}}$\  & $\Delta_{\mbox{orb}}$\  & & $\Delta_{\mbox{scf}}$\  & $\Delta_{\mbox{orb}}$\   \\
\\
\hline
    1  & 6.75E$-$02 &1.67E$+$00 && 3.55E$-$02 & 3.41E$+$00\\
    2  & 1.87E$-$03 & 1.56E$-$01 && 1.19E$-$03 & 2.22E$-$01\\
    3  & 1.12E$-$04 & 5.00E$-$02 && 2.48E$-$05 & 5.92E$-$02\\
    4  & 7.98E$-$06 & 1.30E$-$02 && 2.55E$-$06 & 1.57E$-$02\\
    5  & 5.91E$-$07 & 3.62E$-$03 && 3.11E$-$07 & 6.73E$-$03\\
    6  & 4.44E$-$08 & 9.90E$-$04 && 1.18E$-$08 & 1.95E$-$03\\
    7  & 3.36E$-$09 & 2.73E$-$04 && 2.53E$-$09 & 5.92E$-$04\\
    8  & 2.55E$-$10 & 7.52E$-$05 &&1.38E$-$10 & 2.28E$-$04\\
    9  & 1.93E$-$11 & 2.07E$-$05 && 1.95E$-$11 & 6.18E$-$05\\
   10  & 1.47E$-$12 & 5.72E$-$06 && 1.97E$-$12 & 2.38E$-$05\\
   11  & 1.13E$-$13 & 1.58E$-$06 && 1.59E$-$13 & 8.70E$-$06\\
   12  & 7.43E$-$15 & 4.35E$-$07 && 2.71E$-$14 & 2.17E$-$06\\
   13  & 3.53E$-$15 & 1.20E$-$07 && 3.24E$-$15 & 7.98E$-$07\\
   14  & 2.19E$-$15 & 3.32E$-$08 && 2.93E$-$15 & 2.06E$-$07\\
   16  & 3.41E$-$15 & 2.59E$-$09 && 4.62E$-$16 & 2.76E$-$08\\
   18  & 2.44E$-$16 & 3.21E$-$10 && 0.00E$+$00 & 3.29E$-$09\\
   20  & 0.00E$+$00 & 2.46E$-$10 && 7.70E$-$16 & 2.01E$-$09\\
   25  & 2.44E$-$16 & 4.05E$-$10 && 3.08E$-$16 & 1.47E$-$09\\
   30  & 1.22E$-$16 & 3.71E$-$10 && 1.54E$-$15 & 3.31E$-$09\\
   40  & 1.22E$-$16 & 3.88E$-$10 && 1.69E$-$15 & 1.03E$-$09\\
   50  & 0.00E$+$00 & 2.23E$-$10 && 2.16E$-$15 & 1.97E$-$09\\
   60  & 1.22E$-$16 & 2.05E$-$10 && 0.00E$+$00 & 1.67E$-$09\\
   70  & 0.00E$+$00 & 9.39E$-$11 && 1.54E$-$15 & 1.69E$-$09\\
   75  & 0.00E$+$00 & 9.39E$-$11 && 9.24E$-$16 & 2.01E$-$09\\
\hline
\end{tabular}
\end{table}

In both \Verb!DBSR-HF! and \Verb!GRASP!,  orbitals are improved one at a time until self-consistency is reached.  
 It is interesting to see how the results converge to machine precision in such an iterative process.  Table~\ref{tab:conv} shows how values of the relative change in the total energy and the maximum relative change in the largest change in an orbital in a given iteration, evolve for Be and Rn calculations. Clearly the convergence for Be is faster than Rn but not a lot.  The total energy in both cases is near machine precision in 12 iterations. The orbital parameters still continue to improve but in Rn, never reach the requested precision, {\tt orb\_tol=1.d-10}. In variational methods, the energy is used to determine the wave function. It is the accuracy of the latter that is important when the wave function is used to determine other properties. The limits on accuracy of double precision has never been investigated, to our knowledge, but the present study suggests that more precision is needed for tighter convergence requirements.

%%%%%%%%%%%%%%%%%%%%%%%%%%%%%%%%%%%%%%%%%%%%%%%%%%%%%%%%%%%%%%%%%%%%%%%%%%%%%%%
%
%   SECTION 7. Concluding remarks
%
%%%%%%%%%%%%%%%%%%%%%%%%%%%%%%%%%%%%%%%%%%%%%%%%%%%%%%%%%%%%%%%%%%%%%%%%%%%%%%%
\section{\label{sec7}Concluding remarks}

Our study of the virial theorem has shown how extensively the variational method relies on cancellation in the computation. For the orbital equation the positive kinetic energy and the negative potential energy representing the interaction between an electron and the nucleus are the largest contributors to the total energy, yet they nearly cancel, particularly as the nuclear charge increases.  All calculations for this study were done in double precision of about fifteen (15) significant digits. What is important is the the accuracy of the sum of these two integrals. Possibly these two quantities should be computed in quadruple precision.  It also showed how the residual for the pairs of orbitals needed to be balanced.
Thus simultaneous updates of orbitals may be better able to balance the virial theorem conditions that accurate solutions of the variational problem need to accomplish.  The performance of these methods as implemented in an $\mathtt{SPHF}$~\cite{FroFis:2011} program was better for the NR method when orbital rotation in not used. But most tests focused on light atoms -- what is needed are stable and accurate multiconfiguration methods for systems with several open shells such as the lanthanides and actinides.  The
virial theorem could be a useful tool.

\section*{Ackowledgments}

C. Froese Fischer acknowledges support from the Canada NSERC Discovery Grant 2017-03851. M. Godefroid acknowledges support from the FWO \& FNRS Excellence of Science Programme (EOS-O022818F).
 
\clearpage

\bibliographystyle{pccp}

\begin{thebibliography}{10}

\bibitem{review}
C. Froese Fischer, M. Godefroid, T. Brage, P. J\"onsson, G. Gaigalas,  Advanced multiconfiguration methods for complex atoms: I. Energies and wave functions. J. Phys. B At. Mol. Opt. Phys. {\bf  2016}, {\em 49}, 182004. 

\bibitem{grant}
I. P. Grant, {\em Relativisitic Quantum Theory for Atoms an Molecules}, Springer Science, New York, NY.  {\bf 2007}.

\bibitem{mcdf}
I.P. Grant, B.J. McKenzie, P. H. Norrington, D.F. Mayers, an N. C. Pyper, {\em Comput. Phys. Commun.}. (1980). {\bf 21}, 207-231. 

\bibitem{FroSen:2020a}
C. Froese Fischer and A. Senchuck,
Numerical Procedures for Relativistic Atomic Structure Calculations, {\em Atoms} {\bf 2020} {\em 85}, 1-20.

 \bibitem{dbsr-hf}
O. Zatsarinny and C. Froese Fischer, DBSR -- A B-spline Dirac-Hartree-Fock program,  {\em Comput. Phys. Commun.}, {\bf 2016}, {\em 202}, 287-303.

\bibitem{bspline-advances}
C. Froese Fischer, B-splines in variational atomic structure calculations, {\em Advances in Atomic, Molecular, and Optical Physics}, {\bf 2008}, {\em 55} 235-291.

\bibitem{Goletal:2000a}
H. Goldstein, C. Poole and J. Safko, {\em Classical Mechanics}, Addison Wesley, San Francisco. {\bf 2000}

 \bibitem{finkelstein}
B. Finkelstein, {\em Zs. Phys.} 50, 293, 1928.

 \bibitem{fock}
 V. A. Fock, A comment on the virial relation,  
 {\em Zs. Phys.} {\bf 1930}, {\em 63} 855 (in German). 
\bibitem{fockcoll}
L.D. Faddeev, L.A. Khalfin, and I.V. Komarov, {\it Selected Works -- V. A. Fock -- Quantum Mechanics and Quantum Field Theory}
 ( Chapman \& Hall) 2004. 
 
 \bibitem{Low:59a}
P.-O. L\"owdin,
Scaling problem, virial theorem, and connected relations in quantum mechanics, {\em Journal of Molecular Spectroscopy} {\bf 1959} {\em 3}, 46-66.

\bibitem{Kim:67a}
Y-K Kim,
Relativistic Self-Consistent-Field Theory for Closed-Shell Atoms,
{\em Phys. Rev. A} {\bf 1967} {\em 154}, 17-39.

 \bibitem{matsuoka}
 O.  Matsuoka and T.  Koga,
{\it Relativistic virial theorem for atom}
{\em Theoretical Chemistry Accounts} {\bf 2001} {\em 105}, 473?476. 

\bibitem{ParMoh:92a}
F.A. Parpia and A.K. Mohanty,
Relativistic basis-set calculations for atoms with Fermi nuclei,
{\em Phys. Rev. A} {\bf 1992} {\em 46}, 3735-3745.

\bibitem{Paretal:96a}
F.A. Parpia; C. Froese Fischer and I.P. Grant,
GRASP92: A package for large-scale relativistic atomic structure calculations,
{\em Comput. Phys. Commun.} {\bf 1996}, {\em 94}, 249.

\bibitem{g18}
C. Froese Fischer, G. Gaigalas, P. J\"onsson, J. Biero{\'n},  GRASP2018 -- A Fortran 95 version of the General Relativistic Atomic Structure Package. {\em Comput. Phys. Commun.} {\bf 2019}, {\em 237}, 184?187.

\bibitem{Har:57a}
D.R. Hartree, {\em The calculation of atomic structures}, John Wiley and Sons, New York. {\bf 1957}

\bibitem{FroGuoShe:92}
C. Froese Fischer, W. Guo, and Z. Shen,
Spline Methods for Multiconfiguration  Hartree-Fock Calculations, 
{\em Int'l J. Quantum Chem.}, {\bf 1992} {\em  42}, 849-867.

\bibitem{FroFis:2011}
C. Froese Fischer,  A B-spline Hartree-Fock Program,
{\em Comput. Phys. Commun.} {\bf 2011} 1315-1326.

\end{thebibliography}

  \end{document}